\documentclass[]{spie}  

\newcommand\arcsec{\mbox{$^{\prime\prime}$}}%
\newcommand\degree{\mbox{$^{\circ}$}}%

\usepackage{amsmath,amsfonts,amssymb}
\usepackage{graphicx}
\usepackage[colorlinks=true, allcolors=blue]{hyperref}
\usepackage[colorinlistoftodos]{todonotes}
\usepackage{tablefootnote}
\usepackage{threeparttable}

\title{Modeling and budgeting fiber injection efficiency for the Maunakea spectroscopic explorer (MSE)}

\author[a]{Nicolas Flagey}
\author[a]{Kei Szeto}
\author[b]{Shan Mignot}
\author[a]{Alexis Hill}
\author[c]{Alan McConnachie}
\author[a]{Calum Hervieu}

\affil[a]{Canada-France-Hawaii Telescope Corporation, 65-1238 Mamalahoa Hwy, Kamuela HI 96743, USA}
\affil[b]{GEPI, Observatoire de Paris, PSL Research University, CNRS, Univ. Paris Diderot, Sorbonne Paris Cit\'e, Place Jules Janssen, 92195 Meudon, FRANCE}
\affil[c]{NRC Herzberg, Dominion Astrophysical Observatory, 5071 West Saanich Road, Victoria, British Columbia, Canada}

\authorinfo{Further author information: (Send correspondence to N.F.)\\N.F.: E-mail: flagey@cfht.hawaii.edu}

\begin{document} 
\maketitle

\begin{abstract}
The Maunakea Spectroscopic Explorer (MSE) will each year obtain millions of spectra in the optical to near-infrared, at low ($R\simeq 3,000$) to high ($R\simeq 40,000$) spectral resolution by observing $>$4,000 spectra per pointing via a highly multiplexed fiber-fed system. Key science programs for MSE include black hole reverberation mapping, stellar population analysis of faint galaxies at high redshift, and sub-km/s velocity accuracy for stellar astrophysics.

One key metric of the success of MSE will be its survey speed, i.e. how many spectra of good signal-to-noise ratio will MSE be able to obtain every night and every year. The survey speed is directly linked to the allocation efficiency - how many fibers in the focal surface can be allocated to targets - and to the injection efficiency - what fraction of light from a target can enter the fiber at the focal surface.

In this paper we focus on the injection efficiency and how to optimize it to increase the signal-to-noise ratio of targets observed in sky dominated conditions. The injection efficiency depends on the size of the fiber and requires highly precise, repeatable and stable positioning of the fiber in the focal surface. We present the allocation budget used for Conceptual Design Review and the modeling that allows to estimate the injection efficiency, which is just one part necessary to meet the science requirements on sensitivities.
\end{abstract}


\section{INTRODUCTION}
\label{sec:intro}  

The Maunakea Spectroscopic Explorer (MSE) is a project to upgrade the 3.6-meter telescope and instrumentation of the Canada-France-Hawaii Telescope (CFHT) into a 11.25-meter telescope equipped with fiber-fed spectrographs dedicated to optical and near-infrared (NIR) spectroscopic surveys. The current baseline for MSE is that of a prime focus, 10-meter effective aperture telescope feeding a bank of low and moderate spectral resolution spectrographs (LR, R$\sim$2000-3500 and MR, R$\sim$6000) located on platforms, as well as high spectral resolution spectrographs (HR, R$\sim$20000-40000) located in the more stable pier of the telescope. The 1.5 square degree field of view of MSE will be populated with 3249 fibers of 1\arcsec\ diameter allocated to the LMR spectrographs, and 1083 fibers of 0.8\arcsec\ for the HR spectrographs. The fibers will all be positioned with spines from the Sphinx system, with both LMR and HR fibers being available at all time.

At the previous SPIE Astronomical Telescopes and Instrumentation meeting, the status and progress of the project were detailed in Ref.~\citenum{Murowinski2016} while an overview of the project design was given in Ref.~\citenum{Szeto2016} and the science based requirements were explained in Ref.~\citenum{McConnachie2016}. An update of the project at the end of conceptual design phase is presented this year in Ref.~\citenum{Szeto2018a} with a review of the instrumentation suite in Ref.~\citenum{Szeto2018b}. Other papers related to MSE are focusing on: the summit facility upgrade (Ref.~\citenum{Bauman2016, Bauman2018}), the telescope optical designs for MSE (Ref.~\citenum{Saunders2016}), the telescope structure design (Ref.~\citenum{Murga2018}), the design for the high-resolution (Ref.~\citenum{Zhang2016, Zhang2018}) and the low/moderate-resolution spectrograph (Ref.~\citenum{Caillier2018}, the top end assembly (Ref.~\citenum{Mignot2018, Hill2018b}), the fiber bundle system (Ref.~\citenum{Venn2018, Erickson2018}), the fiber positioners system (Ref.~\citenum{Smedley2018}), the systems budgets architecture and development (Ref.~\citenum{Mignot2016, Hill2018}), the observatory software (Ref.~\citenum{Vermeulen2016}), the spectral calibration (Ref.~\citenum{Flagey2016a, McConnachie2018a}), the throughput optimization (Ref.~\citenum{Flagey2016b, McConnachie2018b}), the observing efficiency (Ref.~\citenum{Flagey2018b}), and the overall operations of the facility (Ref.~\citenum{Flagey2018a}).

In this paper we focus on the injection efficiency (IE) into the fiber at the focal surface, a key element of the overall throughput of the observatory. Given the design choices for MSE, in particular the fiber sizes and the fiber positioner technology, a systems budget was established and a model was developed to estimate the expected IE. The plan of the paper is as follows. In section \ref{sec:key} we define the IE, in section \ref{sec:sys} we detail the budget, and in section \ref{sec:mod} we present our model and how the budget was used to compute the IE.

\section{Key elements of the injection efficiency}
\label{sec:key}  

In this paper, the IE is defined as the fraction of light from a point source that enters a fiber at the focal surface, relative to the amount of light from that target that reaches the focal surface. The IE is thus a component of the overall throughput of the observatory. The IE does not affect the sky light the same way it affects the target's light: the sky surface brightness, assumed spatially flat especially on the scale of the fiber diameter, is only scaled by the size of the fiber to compute the amount of light going through the fiber, while the target produces a point spread function at the focal surface which is truncated by the fiber diameter. Consequently, the throughput of MSE is considered separately from the IE in terms of systems budget. The addition of the noise budget allows to compute the sensitivity of MSE.

The IE is directly related to the diameter of the fiber and to the point-spread function delivered by the telescope and atmosphere, i.e. the image quality of the observatory. From a systems engineering point of view, the IE budget allocates a requirement to the image quality at the focal surface, which is then developed into its own budget. The IE budget focuses on the accuracy with which the fiber and target can be positioned at the focal surface and is detailed in the following sections. Hereafter we briefly discuss the fiber sizes and the image quality.

In the current configuration of MSE, the fiber diameter is set at 1.0\arcsec\ for the Low-Moderate Resolution (LMR) spectrographs and 0.8\arcsec\ for the High Resolution (HR) spectrographs. The fiber size has to be as large as possible to maximize the flux from the target reaching the detectors in the spectrographs, but as small as possible to limit the amount of sky light reaching those same detectors, especially for sky-limited observations. If the image quality at the focal surface is good, the fiber sizes can be reduced to optimize the signal-to-noise. If the image quality is bad, then the fiber sizes should be increased. Taking into account the natural seeing distribution at the CFHT site, a sweet spot is found near the currently selected sizes. We are currently refining these sizes given the completed systems budgets and expect that the diameters will not change by more than 0.1\arcsec. For the LMR fibers, which will most likely be allocated to galaxies, the angular size of the targets will be folded into the simulations at a later stage.

The details about the point-spread function at the focal surface are given in the Image Quality budget\footnote{Project internal document available upon request}, which is dominated by the natural seeing, and thus emphasizes how critical it is to choose a site known for its exquisite image quality. The summit of Maunakea is definitely one of the best on the planet. The Thirty-Meter-Telescpe (TMT) reported a median seeing of 0.5\arcsec\ at a wavelength of 500nm and at 60m above ground for their Hawaiian site and their Chilean site\footnote{\url{https://www.tmt.org/download/MediaFile/126/original}}. At the Eastern ridge on the summit of Maunakea, where the CFHT is located and where MSE is planned to be built, the median image quality derived from 10 years of measurements by the MKAM-DIMM is 0.62\arcsec. This value is then converted into an IQ$_{500}$, as follows:

\begin{itemize}

\item MKAM-DIMM measurements (median value of 0.62\arcsec) obtained at various airmass (AM) are converted to Zenith values assuming $IQ_{AM=1.0} = IQ_{AM} / AM^{3/5}$, where $IQ_{AM=1.0}$ is the image quality at Zenith.

\item The effect of the ground-layer, studied by Ref.~\citenum{Gagne2011}, is removed assuming a contribution of 0.289\arcsec\ between 7~m and 22~m elevation (the difference between MKAM-DIMM and the observatory): $IQ_{no GL} = (IQ_{AM=1.0}^{5/3} - 0.289^{5/3})^{3/5}$. This leads to a median value of 0.51\arcsec.

\item the value of $r_0$ is computed at 500~nm using Ref.~\citenum{Tokovinin2002}: $r_0 = 0.98 * 500e-9 / ([IQ_{no GL} / 180] * [\pi / 3600])$ and used to derive values for IQ$_{500}$  assuming $L_0 = 30$~m: $IQ_{500} = IQ_{no GL} * \sqrt(1 - 2.183 * (r_0 / 30)^{0.356})$. This then leads to a median value of 0.41\arcsec.

\end{itemize}

In the IE budget we then assume this median value of 0.41\arcsec\ for $IQ_{500}$. We then assume additional IQ allocations of 0.25\arcsec\ FWHM from all other contributors in the IQ budget which, once added in power $5/3$, lead to an image quality of 0.5\arcsec, at 500~nm, at the focal surface of MSE (or 1.08\arcsec\ diameter in the 80\% encircled energy metric).

\section{Systems budget}
\label{sec:sys}  

We break down the systems budget for the IE as follows. The first step corresponds to the ideal theoretical model for the MSE system without any defects. Then, we take into account the contributions from the as-delivered positioners and fiber bundle systems by the suppliers. We then incorporate the contributions incurred during the assembly integration and verification (AIV) phase. Finally, the most complex step is encompassing all effects that occur during operations. For operations consideration, the effects are groups sequentially starting with setup, then acquisition, and eventually exposure. The setup is the first step in the observing process and after the acquisition step, the system is ready for the exposure. In the budget, the exposure step is further organized according to effects associated with functionalities including the telescope motion in guiding, differential atmospheric refraction after correction by the atmospheric dispersion corrector (ADC), plate scale variation, and sources of fiber tip motions during exposure.

For each contributor, we specify whether it is a lateral effect or a longitudinal effect. A lateral effect remains within the focal surface while a longitudinal effect is perpendicular to the focal surface and can usually be related to defocus. We also specify if the effect is constant during an observation, if it depends on the positioner, if it is varying smoothly and slowly or randomly and quickly during an observation. These precisions are important for the way the effects are accounted for in the IE calculations (see section \ref{sec:mod}).

\subsection{Theoretical model}
\label{sec:sys_theor}  

The chromatic aberrations lead to different wavelengths not being centered at the same position at the focal surface, which will decrease the IE at some wavelengths more than others, even after correction by the ADC. Positioning the fibers has to be optimized to maximize survey efficiency according to some metric, depending on the wavelength coverage and science priority. These aberrations are inherent to the optical design of MSE.

Using the Zemax optical design of the telescope, we can measure the distance between the chief rays of different wavelengths in any configuration (i.e. zenith distance, position within field of view). From this analysis, we derived that the maximum distances between chief rays of different wavelengths is 88 microns at 50\degree\ Zenith distance. Limiting the analysis to zenith distance less than 30 degrees, where most of the observations are expected to take place, the maximum lateral chromatic displacement is 41 microns. Adding a 10\% margin brings the allocation to 45 microns.

\begin{figure}[h]
\centering
\caption{\label{fig:tea} Details of the prime focus design of MSE telescope with locations of most critical subsystems. The fiber bundles, not shown here, will run along the telescope structure, from the top end assembly to the spectrographs on the platform and in the pier (not shown).}
\includegraphics[width=0.475\linewidth]{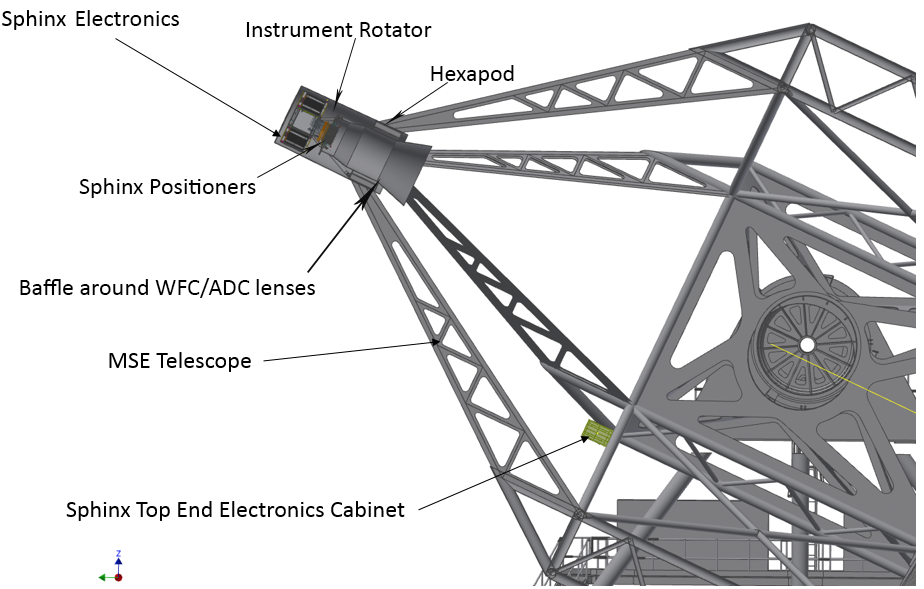}
\includegraphics[width=0.475\linewidth]{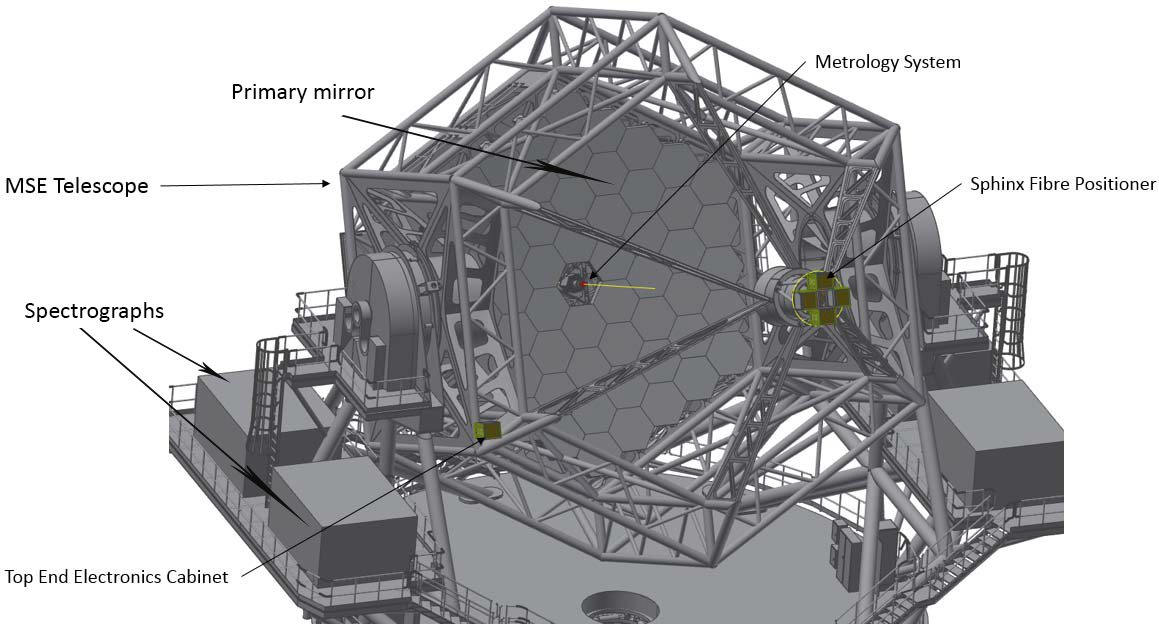}
\end{figure}

\subsection{As delivered}
\label{sec:sys_deliv}  

We consider here only the combined subsystems of the positioners and fibers bundles, after they have been assembled together at the supplier but before they are brought to the telescope to be assembled and integrated with other top end subsystems (see Figure \ref{fig:tea}). Ideally, the fiber tips are perfectly aligned with the expected focal surface delivered by the telescope. In reality, the positioners are longitudinally scattered around their required locations and the fiber tips, once installed in the positioners, contribute to additional scatter with respect to the focal surface.

The Sphinx positioner system will provide 99.5\% (goal of 100\%) of all spines, when located at their home position, parfocal to $\pm$25 microns. Once the fibers are inserted in the positioners, the scatter of their tips positions with respect to the focal surface will increase beyond 25 microns. This will be a random effect to which we allocate an additional 5 microns, based on the tolerance proposed by the NRC-HAA design team.

\subsection{AIV}
\label{sec:sys_aiv}  

During the AIV phase, the positioners and fibers bundles subsystems are integrated to the telescope. After the integration to the telescope, residual misalignment of the interface plane will be corrected by shims such that the subsystems are mounted at the expected focal surface location. We consider here a residual displacement, uncorrected by these shims or any other way, including during operations.

The longitudinal allocation of 50 microns is based on the positional accuracy achievable by a well-calibrated laser tracker system which is part of the telescope Global Metrology System (GMS) and include both focus and tip/tilt errors. A portion of this alignment error can be compensated by the Prime Focus Hexapod System with feedback from the Telescope Optical Feedback System (TOFS). This allocation will be revised in the future pending the completion of the TOFS conceptual design.

\subsection{Operations}
\label{sec:sys_ops}  

The use of lookup tables to account for deformation of the structure due to gravity and thermal effects will not be perfect and lead to residual errors (section \ref{sec:sys_ops_fmod}). This affects the focus model only, since the pointing model errors will be corrected by guiding. These contributions do not vary during an observation but on longer timescales. The relative fiber positioning error is a lateral effect (section \ref{sec:sys_ops_poss}) while the tilt inherent to the spine design induces a longitudinal defocus (section \ref{sec:sys_ops_tilt}). Those contributions are fixed during an exposure, but vary from positioner to positioner. The telescope motion due to finite guiding accuracy leads to lateral errors (section \ref{sec:sys_ops_guid}). This contribution varies during an observation rapidly/randomly. The plate scale variations (section \ref{sec:sys_ops_ps}) and differential atmospheric refraction (section \ref{sec:sys_ops_adr}) create lateral errors. These contributions vary during an observation, though slowly and monotonically. The motion of fiber tips during the exposures (i.e. on short time scales) generates lateral and longitudinal errors (section \ref{sec:sys_ops_oth}). These contributions vary during an observation, some rapidly/randomly and some slowly and monotonically.

\subsubsection{Focus model}
\label{sec:sys_ops_fmod}  

A focus model will be established over time for MSE. Lookup tables will be built to preserve the “best focus” position based on actual zenith angle (i.e. gravity) and temperature conditions for any given observation. However, the focus model and the exact conditions during observations, especially temperature, will not be known with infinite accuracy, leading to residual errors in the focus of MSE. 

The flexure of the telescope structure due to thermal and gravitational effects results in (lateral) pointing and (longitudinal) best-focus variations of the telescope. These variations are compensated by lookup table adjustments, i.e. pointing and focus models, when configuring for the upcoming observation.

We allocate 10 microns maximum to the longitudinal error, which will vary from one observation to another, but will remain constant during an observation. The 10 micron allocation is based on operation experience with the current CFHT focus model.

Pointing variations are eliminated once the target acquisition is established and telescope guiding commences (see section \ref{sec:sys_ops_guid}).

Moreover, telescope flexure over timescales similar to the duration of an observation will introduce additional lateral errors. The PFHS and TOFS provide corrections for those effects during an observation. Their functionalities are outlined in sections \ref{sec:sys_ops_oth} and \ref{sec:sys_ops_guid}, respectively.

\subsubsection{Positioning accuracy}
\label{sec:sys_ops_poss}  

The errors in the lateral fiber positioning relative to the actual targets come from at least three sources:
\begin{itemize}

\item inaccurate target coordinates because the astrometry was not accurately determined, or because the conversion from sky to MSE coordinates is not perfect. We expect targets coordinates to be provided with a precision a small fraction of 1\arcsec, especially in the post-GAIA and large imaging surveys era. However, fainter galaxies will suffer from a significantly larger uncertainty: a magnitude 24 galaxy is about 40 times fainter than a magnitude 20 star, leading to an SNR about 6 times lower for the same integration time. The allocated error of 0.1\arcsec\ corresponds to 10 microns at the focal surface. 

\item residual errors in mapping of the sky coordinates to focal surface coordinates. This may be because the fiducial positions in the focal surface and in the acquisition/guide camera system are not precisely known, because our model of the focal surface is inaccurate, and/or because of inaccuracy of the acquisition/guide camera system. The sky coordinates to focal surface mapping error depends on the precision to which we determine the fiducial locations on the guiding cameras and in the focal plane, the inaccuracy of the mapping model of the focal surface, and the accuracy of the acquisition/guide cameras (which is small but not infinitely small). We expect that the contributions from SIP.TOFS will be small after calibration to within a fraction of a pixel, so typically about a micron. According to the AAO-Sphinx design team, the mapping error after calibration is attributed to the metrology system and is allocated 4 microns, twice the contribution in closed-loop accuracy.

\item inaccuracy in the FPMS-to-PosS positional feedback loop, which iteratively moves the positioners to their targeted positions in the focal surface with a finite precision. The positioners and metrology combined system closed-loop accuracy allocations are those demonstrated by the AAO/Sphinx design team at 6 microns rms. The error remains constant during an observation but varies from positioner to positioner.

\end{itemize}

\subsubsection{Tilt induced defocus}
\label{sec:sys_ops_tilt}  

Since the patrol regions of individual positioners do not conform to the “global” focal surface, some additional defocus is expected depending on the actual position of the fiber within each patrol region, i.e. the spine's tilt. Given the patrol region (1.24 times the pitch of 7.77 mm) and the length of the spines, we estimated the longitudinal defocus displacement due to the tilt of the spines: with 250 mm spines the maximum longitudinal amplitude due to tilt is 186 microns, while it is 156 microns for the longer 300 mm spines. We have opted for the 300 mm spines and allocated a maximum longitudinal error of $\pm$80 microns. We assume here that the average tilt will position the fiber tip at the focal surface.

\subsubsection{Guiding}
\label{sec:sys_ops_guid}  

During observations, many subsystems operate in parallel to maintain acquisition: the guider performs accurate position adjustments of the telescope, the instrument rotator follows the rotation of the field, etc. The precision with which these motions are performed will influence the IE as they will move the targets and/or the fibers in the field of view.

The CFHT guiding error is about 0.1\arcsec\ rms, which corresponds to 10 microns rms at the focal surface. It is similar for the Keck telescopes. We allocate 7 microns rms each, added in quadrature, to the telescope mount and the telescope optical feedback system.

The instrument rotator rotation rate is expected to be accurate to 5 microns rms at the edge of the field, which corresponds to 3.5\arcsec\ rms angular positioning accuracy.

\subsubsection{Plate scale variations}
\label{sec:sys_ops_ps}  

The accurate positioning of the fiber tips in the focal plane relies on a well-known plate-scale. If the scale varies during an observation, the position of the targets in the field of view will drift and the IE will decrease. This is a budget margin for additional optical effects that are not accounted for in the IE budget allocation, e.g. mirror cell deformation.

A change of the plate-scale by 1\% will “move” a target at the edge of the field of view by 1\% of the field of view radius or about 3~mm. This is not tolerable but trackable and correctable by optical feedback (acquisition/guide cameras and metrology systems). We allocate a drift of 10 microns maximum (i.e. at the edge of the field) which corresponds to a  relative plate-scale variation of 0.003\%.

\subsubsection{Atmospheric differential refraction}
\label{sec:sys_ops_adr}  

The targets will drift away from each other during an observation because of the atmospheric differential refraction (ADR). Therefore an optimal positioner configuration setup to maximize the IE during the whole observation will be adopted.

The optical design of MSE limits the ADR drift to 78 microns between 0 to 60 degree zenith, with most of the drift (40 microns) occurring between 50 to 60 degree zenith. Assuming most observations will happen at 30 degree zenith and will be shorter than one hour, which limits the change in zenith angle to 15 degrees, and assuming the fibers will be positioned at the “average” expected position, we allocate +/- 15 microns maximum to the ADC-corrected ADR residual.

As a future enhancement, the Sphinx positioners can potentially operate in open-loop to follow the targets and mitigate the effect of the ADR. If open-loop fiber positioning is implemented, the lateral motion during an observation due to ADR can be reduced to 6 microns rms for steps size up to 40 microns.

\subsubsection{Other motions of the fibers}
\label{sec:sys_ops_oth}  

During an observation, the fiber tips may be displaced, both laterally and longitudinally, due to external effects, in both long and short timescale variations, from gravity, temperature, hexapod correction, instrument rotator tilt, and other vibration sources such as wind, plumbing and rotating mechanisms, etc.

The thermal loads affect individual positioners during observations due to expansion/contraction of the positioners support structure. Expansion/contraction of the positioners support structure leads to the lateral displacement of individual positioners by different amounts according to their radial distance and temperature gradient. There is also a longitudinal displacement through the expansion/contraction of the thickness of the support structure which is uniform across the field of view. We expect the expansion of the positioners to be limited to about 14 microns per hour laterally, and 1.5 microns longitudinally. Given the heat capacity of steel (500~J/kg/K), it requires 100~kJ to heat 100~kg of steel (mass of the base plate and support structure) by two degrees (the maximum temperature change observed at CFHT in one hour). If the air temperature is at $\pm$2 degrees, natural convection provides about 60~W per unit surface of exchange. Given the size of the field of view, and assuming the positioners support stucturecan be heated from the top and bottom, the exchange surface is about 0.5~m$^2$. In sum, the steel mass would warm up by two degrees in slightly less than one hour. Then, such temperature increase of a 0.59~m by 60~mm piece of steel leads to an expansion of 14 microns laterally and 1.5 micron longitudinally. We recognize our assumptions are very conservative.

The gravity loads affect the entire positioners and so the lateral displacement can be corrected by the acquisition/guide camera system. The longitudinal displacement due to sag cannot however be corrected. The design from the AAO/Sphinx predicts a 4 micron sag at zenith. We conservatively allocate only 1 micron maximum to the gravity {\em variation} on the positioners during observation. This variation is expected to be smooth, following the variation of the gravity vector while guiding.

The instrument rotator axis is assumed to tilt very slowly, over a full revolution, which is typical of rotary bearings, and thus produce a smooth variation of longitudinal displacement. We allocate 30 microns maximum longitudinal displacement at the edge of the field, which corresponds to a tilt of about 20\arcsec\ of the instrument rotator structure. Given the instrument rotator may rotate by 180º in one hour, the maximum amplitude of defocus is thus expected to be up to $\pm$30 microns, at the edge of the field, and negligible near the optical axis.

The hexapod is utilized to correct the overall focus during observation and incurs additional errors from the focus model, which have been discussed previously. The hexapod has a limited precision in its movement, to which we allocate 5 microns maximum. We envision the hexapod correction will occur several times during an observation, maybe about once every 15 minutes.

Lastly, we allocate a maximum of 1 micron to the lateral displacement due to vibration of the positioners via the telescope structure as a placeholder pending future dynamic analysis. Inertia effects on the positioners due to telescope and instrument rotator acceleration are expected to be negligible based on shake-test results presented by the AAO/Sphinx design team.

\section{Modeling}
\label{sec:mod}  

We use the budget presented above to compute the actual IE as follows. We begin with the 2-dimensional point-spread function (PSF) from the 60 individual segments in a given configuration (zenith distance, defocus, field position, wavelength) as computed in Zemax. We then combine all 60 segment PSFs for each configuration and combine all wavelengths into a PSF cube. We then convolve this PSF cube with the image quality of the observatory. This thus provides the polychromatic ideal PSF delivered by the telescope. We finally obtain IE curves by integrating the PSF within a disk whose diameter represents the fiber, and whose positions are simulating the accuracy with which the system places fibers in the focal surface.

\begin{table}
\caption{\label{tab:param} Summary of the parameters used in the model}
\begin{threeparttable}
\begin{tabular}{| l | l |}
\hline
{\bf Parameter} & {\bf Value} \\
\hline
\multicolumn{2}{| l |}{Zemax} \\
\hline
Segment size\tnote{a} & 1.44 m \\
Sampling at focal plane\tnote{b} & 0.3~$\mu$m per pixel \\
Image size\tnote{c} & 512 by 512 pixels \\
\hline
\multicolumn{2}{| l |}{Configurations} \\
\hline
Defocus\tnote{d} & $\pm$100~$\mu$m, $\pm$25~$\mu$m, and at focus \\
Zenith distance\tnote{e} & 0, 30, 50, and 60\degree \\
Field positions\tnote{f} & on axis, X +0.530\degree, +0.650\degree, and +0.750\degree, Y $\pm$0.375\degree, $\pm$0.530\degree, $\pm$0.650\degree, ±0.750\degree \\
Wavelengths\tnote{e} & 360, 370, 400, 445, 551, 658, 806, 1000, 1214, 1477, and 1784 nm \\
\hline
\multicolumn{2}{| l |}{Focal surface} \\
\hline
Focal surface radius & 0.76\degree \\
Plate scale\tnote{g} & 106.7~$\mu$m/\arcsec \\
\hline
\multicolumn{2}{| l |}{Fibers and positioners} \\
\hline
Fiber diameter & 0.8″ for HR and 1.0″ for LMR \\
Positioners pitch & 7.77 mm \\
\hline
\end{tabular}
\begin{tablenotes}\footnotesize
\item[a] Corner to corner
\item[b] Allows to properly sample effects on the IE from all lateral contributors
\item[c] This choice allows to use “small” files (512 by 512 pixels) while covering the very large majority of the flux (at least 87\%, and up to 99\%, depending on the configuration)
\item[d] “At focus” is the polychromatic best focus using the wavelength weighting scale
\item[e] Those are the values in the Zemax model
\item[f] Coordinate system is shown in Figure \ref{fig:seg_vign}
\item[g] Average value used
\item[h] From Sphinx CoDR.
\end{tablenotes}
\end{threeparttable}
\end{table}

\subsection{Primary mirror segments monochromatic PSF}

\begin{figure}
\includegraphics[]{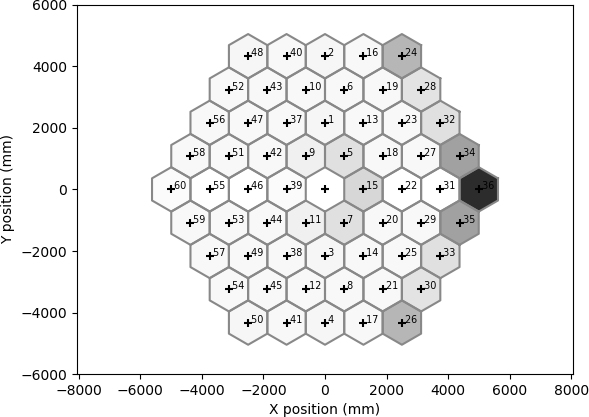}
\caption{\label{fig:seg_vign} Positions and orientation of the 60 segments in the primary mirror of MSE. There is no central segment. The vignetting of each segment is indicated by the greyscale (the more vignetting, the darker) for the field position 0.53$\degree$ from the optical axis, along the X-axis. The segments are numbered.}
\end{figure}

The first step in the process of computing the IE at the entrance of the fibers is to obtain an accurate representation of the PSF as delivered by the optical design only. This has been achieved in Zemax, using version “6u-2” of the optical design proposed for MSE. Because the primary mirror is segmented, a mask had to be generated in Zemax, using a User Defined Aperture (UDA) on Surface 8 (primary mirror) in the design. The UDA file defines a hexagon aperture thanks to the command {\verb POL 0 0 720 6 0}. The aperture central position was then modified via a script. Figure \ref{fig:seg_vign} shows the positions of the segments and their orientation.

\begin{figure}
\includegraphics[width=\linewidth]{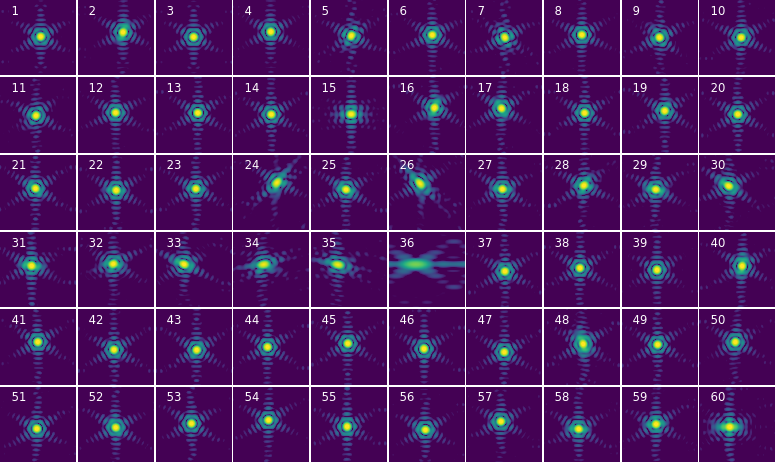}
\caption{\label{fig:seg_psf} Point spread function of the 60 segments in the primary mirror of MSE for a configuration at 30$\degree$ from Zenith, a field position 0.53$\degree$ from the optical axis, along the X-axis, and at 551~nm. A logarithmic scale was used. Note the peculiar shape of segment \#36, the most vignetted of all in that configuration (see Figure \ref{fig:seg_vign}). Segments 24, 26, 34, and 35 are also distorted, though to a lesser extent. The segments numbers and X- and Y-axis are shown on Figure \ref{fig:seg_vign}.}
\end{figure}

The PSF at the focal surface was then computed in Zemax using the Fast Fourier Transform PSF tool (FPS tool), for each configuration (i.e. Zenith distance, wavelength, defocus, and field position) listed in Table \ref{tab:param}. The FPS tool was used to generate images of 153.6~$\mu$m by 153.6~$\mu$m with a 0.3 micron sampling in order to properly sample the most peaked PSF and cover enough of the most spread out, while keeping the required disk space to a reasonable amount. The PSF for each of the 60 individual segments is shown in a given configuration in Figure \ref{fig:seg_psf}.

\subsection{Primary mirror monochromatic PSF}

The Footprint Diagram (FD) tool in Zemax was then used to compute the vignetting/obscuration on each segment, at each zenith distance, and for each field position.

The FD tool basically provides a “percentage of rays through” (PoRT), i.e. the fraction of rays reflected by one segment of the primary mirror and that reach the focal plane. Because those rays are distributed over the entire primary mirror in Zemax, the percentage for each segment is a number smaller than 1.35\%, depending on the segment and the field position (see Figure \ref{fig:seg_vign}). The maximum value, for the 60 segments combined, is found at Zenith, for a target on the optical axis, where the primary mirror collects 81\% of the "rays" simulated in Zemax (60 times 1.35\%). The 81\% corresponds to the fractional surface of 60 hexagonal segments of 1.44m corner-to-corner over a circle of 11.25m, which is the actual aperture size used in Zemax. The 81\% number is thus used as the unvignetted reference value. This number decreases to 71\% in the worst case, at the edge of the field, which means that vignetting at the edge of the field corresponds to 14\% at most ($81/71 = 1.14$).

The PoRT is then used to normalize the 2-dimension PSFs of each segment, after scaling it by 1/0.81. The PSF from all segments in a given configuration are then stacked to generate the PSF of the telescope (M1 + WFC/ADC). In this process, we verify that the coordinates of the chief ray are the same for all segments.

\begin{figure}
\includegraphics[width=\linewidth]{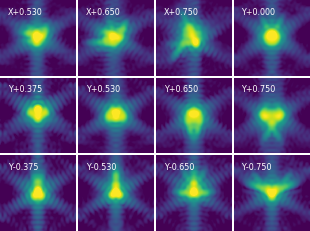}
\caption{\label{fig:m1_psf} Point spread function of the primary mirror of MSE for a configuration at 30$\degree$ from Zenith and at 551~nm. All field positions are shown. A logarithmic scale was used.}
\end{figure}

On axis, at Zenith, the FWHM of the PSF delivered by the optics is about 0.14\arcsec. At the edge of the field of view, it reaches values of 0.14\arcsec\ by 0.29\arcsec. At 50\degree\ from Zenith, the FWHM is slightly better on axis at 0.11\arcsec\ by 0.14\arcsec, but worse at the edge of the field with values that can reach 0.31\arcsec\ in some directions. As a comparison, the diffraction limit of an aperture of 1.44~m is 0.1″\arcsec\ at 550 nm.

\subsection{Primary mirror polychromatic PSF}

The next step consists of combining the monochromatic PSFs from the optical design into a polychromatic PSF, i.e. a spectral cube with 11 frames, one for each wavelength. Since the monochromatic PSFs were obtained in identical conditions (e.g. sampling), the only thing to do when combining them into a cube is to verify that they are properly aligned, using the coordinates of the chief ray, in mm and pixel, as provided by Zemax.

\begin{figure}
\includegraphics[width=.75\linewidth]{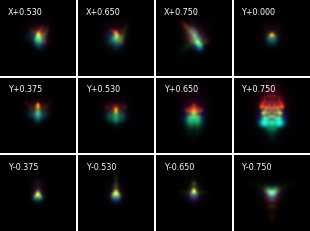}
\caption{\label{fig:m1_polypsf} Polychromatic point spread function of the primary mirror of MSE for all field positions at 30$\degree$ from Zenith. A linear scale was used. The 11 wavelengths were artificially coded into regularly spread values of hues in the HSV (hue, saturation, value) system.}
\end{figure}

\begin{figure}
\includegraphics[width=.75\linewidth]{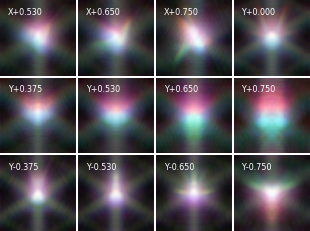}
\caption{\label{fig:m1_polypsf_log} Same as Figure \ref{fig:m1_polypsf} with a logarithmic scale.}
\end{figure}

\subsection{Image quality at the focal surface}

Given the polychromatic PSFs, which represent the polychromatic image quality (IQ) of the optical design alone in different configurations, we can then add the effects that degrade the IQ. Those are described in our IQ Systems Budget. The main contributor is the natural seeing, which has a median value of 0.41\arcsec\ FWHM, but taking into account all other contributors, we allocate a median value of 0.5\arcsec\ FWHM at 550~nm to the entire IQ budget at the focal surface of MSE (see section \ref{sec:key}).

In the IE model, all these effects are accounted for by convolving the optical design PSF with a single Moffat profile ($\beta=3$) of a given FWHM at 550 nm to account for the variations of the natural seeing. The FWHM of the Moffat at all wavelengths other than 550 nm is assumed to vary as a power law of index $-1/5$. The 2-dimensional Moffat kernel is then used to convolve each monochromatic frame of each polychromatic PSF cubes of the optical design. Because the size of the Moffat kernel can exceed that of the PSF, we pad with zeros the outside regions of the PSF spectral cube before we perform the convolution.

\subsection{Optimal lateral position of the fiber}

Once the PSF delivered by the optical design has been degraded to account for natural seeing and other contributors, the optimal position of the fiber in each configuration can be defined. It is derived using the spectral cubes created previously and finding the position, with the 0.3 microns sampling from the Zemax files, where the IE will be the best. To do this, we identify the peaks of the PSF at each wavelength, and define a region that encompasses all these peaks. This region typically spans between a few microns to a few tens of microns, depending on the amplitude of the chromatic aberration.

The IE curve is measured at each position within the region as the fraction of flux going through the fiber diameter and a weighted average of each IE curve is then computed, using the weights from Table 4 and Table 5. The optimal position is that which provides the highest weighted average IE. This is done for each ZD, field position, defocus, natural seeing, and spectrograph (i.e. weights and fiber radius of 0.8” for HR and 1.0” for LMR).

The LR weights are those used to optimize the Zemax optical design. Because the wavelengths used in our analysis are not exactly those used in the optical design, an interpolation was performed, leading to the weights listed in the following tables. The MR and HR weights were fixed by the Science Team and interpolated onto the same wavelength as the LR weights.

\begin{table}
\caption{Weights used in the finding of the optimal position of the fiber at the focal plane.}
\begin{tabular}{c | c c c c c c c c c c c}
\hline
Wavelength     & 360 & 370 &  400 &  445 &  551 &  658 &  806 & 1000 & 1214 & 1477 & 1784 \\
\hline
Weight (LR)    &   1 &   4 & 3.67 & 3.17 & 2.01 & 3.08 & 3.55 &    2 & 3.43 & 2.94 &  1.1 \\
\hline
Weight (MR/HR) &   1 &   3 & 2.83 & 2.58 & 1.99 & 1.46 & 0.78 &    - &    - &    - &    - \\
\hline
\end{tabular}
\end{table}

\begin{figure}
\includegraphics[width=0.45\linewidth]{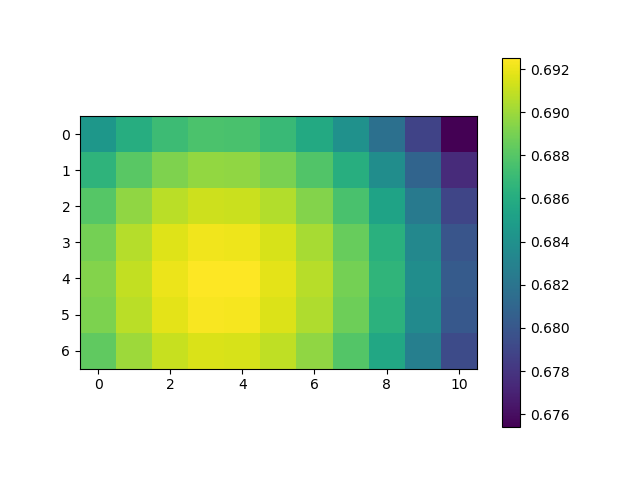}
\includegraphics[width=0.45\linewidth]{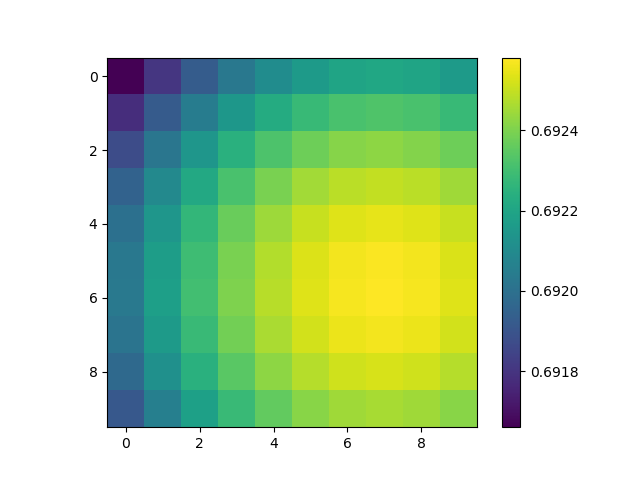}
\caption{\label{fig:optim} Illustration of the method used to find the optimal position for a fiber diameter of 1.0\arcsec, in a configuration at 30$\degree$ from Zenith, a field position 0.53$\degree$ from the optical axis, along the X-axis, a Moffat of 0.5\arcsec\ FWHM, and the weights for the LR spectrograph. A coarse search with 1.5 microns steps is performed first in a small region that surrounds all the monochromatic peaks of emission (left) before a finer search with 0.3 microns steps is performed on a smaller region (right). The colorbar indicates the range of values for the weighted average IE.}
\end{figure}

\subsection{Injection Efficiency curves}

\begin{figure}
\centering
\includegraphics[width=0.75\linewidth]{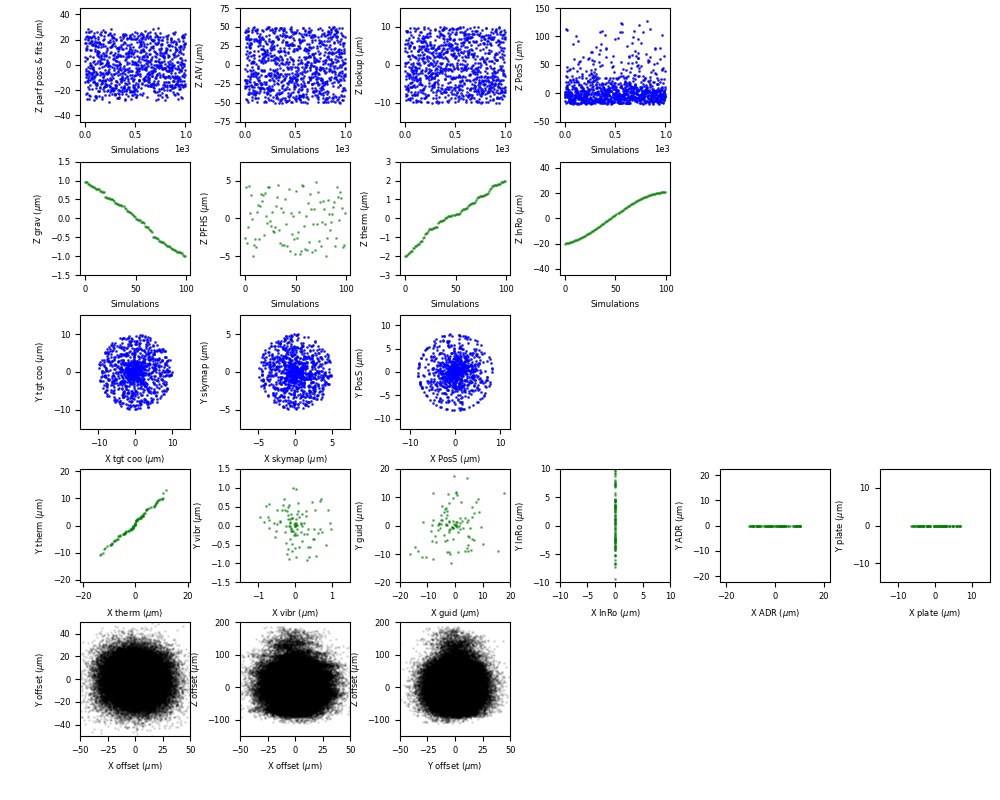}
\caption{\label{fig:simu} Simulations of the various contributors to the IE budget. In these figures, a much larger number of simulations (1000) and samples (100) were used to better show the typical distribution for each contributors. On the left, one can see the typical distribution in physical coordinates (XYZ) while on the right, histograms are shown. The first two rows show offsets in the Z direction while the next two rows show offsets in the XY directions. The last row show the total offsets in each direction. In blue are the contributors that do not vary during an observations while in green are those that do vary. For the latter, we only show the samples associated to one simulation.}
\end{figure}

\begin{figure}
\centering
\includegraphics[width=0.75\linewidth]{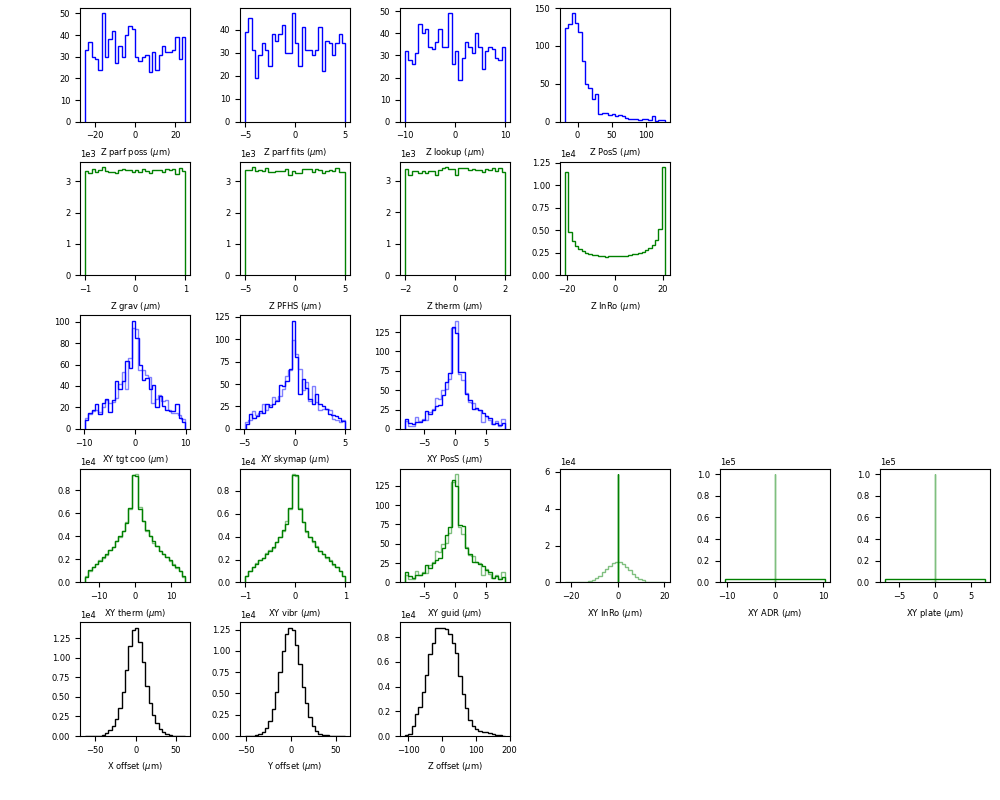}
\caption{\label{fig:simu2} Histograms for the values shown in Figure \ref{fig:simu}.}
\end{figure}

Once the optimal positions have been determined for each configuration, we simulate the positioning errors and their impact on the IE. If no error were made, the fiber would be positioned at the optimal position and the IE would be that which maximize the weighted IE. To simulate the positioning errors, we randomly select 50 values for each contributor to the errors that vary from positioner to positioner but do not vary during an observation. For each of those 50 simulations, we then randomly sampled 25 values for each contributor to the errors that vary during an observation (see section \ref{sec:sys}). Hereafter we describe how each contributor was simulated (see Figures \ref{fig:simu} and \ref{fig:simu2} for details). First, we look at the contributors whose effect is longitudinal and remains constant during an observation.

\begin{description}
\item[Parfocality for positioners (25 $\mu$m) and fibers (5 $\mu$m):] values are randomly picked in a uniform distribution $U(-25, 25)$ for the positioners, and $U(-5, 5)$ for the fibers.
\item[AIV misalignment (50 $\mu$m):] values are randomly picked in a uniform distribution $U(-50, 50)$.
\item[Lookup table residual (10 $\mu$m):] values are randomly picked in a uniform distribution $U(-10, 10)$.
\item[Positioner defocus due to tilt:] values are randomly picked in the distribution of tilts derived from an analysis of the allocation efficiency for the spine design on real fields of view and converted into a defocus assuming a 300~mm spine length.
\end{description}

Then, we look at the contributors whose effect is lateral and remains constant during an observation.
\begin{description}
\item[Target coordinates (10 $\mu$m):] values are randomly picked in a uniform distribution $U(0, 10)$ for the radial error and values are picked in a uniform distribution $U(0, 2\pi)$ for the angular error.
\item[Sky to focal surface mapping (5 $\mu$m):] values are randomly picked in a uniform distribution $U(0, 5)$ for the radial error and values are picked in a uniform distribution $U(0, 2\pi)$ for the angular error.
\item[Positioner accuracy (4.5 $\mu$m):] values are randomly picked in normal distribution $\mathcal{N}(0, 4.5)$ for the radial error with all values above 8.1 set at 8.1 and angular values are picked in a uniform distribution $U(0, 2\pi)$ for the angular error.
\end{description}

Then, we look at the contributors whose effect is longitudinal and varies during an observation.
\begin{description}
\item[Gravity loads (1 $\mu$m):] values are randomly picked in a uniform distribution $U(-1, 1)$ and sorted either increasingly or decreasingly to simulate the smooth variation of that contributor. 
\item[Longitudinal thermal loads (2 $\mu$m):] values are randomly picked in a uniform distribution $U(-2, 2)$ and sorted either increasingly or decreasingly to simulate the smooth variation of that contributor.
\item[Hexapod correction (5 $\mu$m):] values are randomly picked in a uniform distribution $U(-5, 5)$.
\item[Instrument rotator tilt (30 $\mu$m):] a random value is picked in a uniform distribution $U(0, 2\pi)$ to decide the starting angle of the instrument rotator, and a random value is picked in the distribution of instrument rotator's range for a 1-hour observation derived from an analysis at the location of CFHT, with a keyhole at 900\arcsec/s. The effect is scaled by the distance to the optical axis.
\end{description}

Then, we look at the contributors whose effect is lateral and varies during an observation.
\begin{description}
\item[Lateral thermal loads (14 $\mu$m):] values are randomly picked in a uniform distribution $U(-14, 14)$ and sorted either increasingly or decreasingly to simulate the smooth variation of that contributor.
\item[Guiding error (10 $\mu$m):] values are randomly picked in normal distribution $\mathcal{N}(0, 10)$ for the radial error with all values above 8.1 set at 8.1 and values are picked in a uniform distribution $U(0, 2\pi)$ for the angular error.
\item[Vibrations (1 $\mu$m):] values are randomly picked in normal distribution $\mathcal{N}(0, 1)$ for the radial error and values are picked in a uniform distribution $U(0, 2\pi)$ for the angular error.
\item[Instrument rotator accuracy (5 $\mu$m):] values are randomly picked in normal distribution $\mathcal{N}(0, 5)$ at the edge of the field, and hence scaled by the distance to the optical axis. The values are then spread along an arc (i.e. a fixed radius) to simulate the effect of the instrument rotator tracking inaccuracy.
\item[Plate scale variations (10 $\mu$m):] values are randomly picked in a uniform distribution $U(-10, 10)$ and sorted either increasingly or decreasingly to simulate the smooth variation of that contributor. The values are then scaled by the distance to the optical axis and aligned along a radius.
\item[ADR residuals (15 $\mu$m):] values are randomly picked in a uniform distribution $U(-15, 15)$ and sorted to simulate the smooth variation of that contributor. The values are then scaled by the distance to the optical axis and aligned along a radius. According to the optical design report, we assume the effect is centripetal if along the Y axis, and centrifugal along the X axis.
\end{description}

We thus end up with 50 simulations with 25 samples each, and measure the injection efficiency for each sample by integrating the flux of the PSF within the fiber radius moved by the total amount of offset in the X, Y, and Z directions. We do this for each configuration (i.e. Zenith distance, field position, fiber diameter, seeing). To account for the defocus, we interpolate for the simulated defocus value from the five model values ($\pm$100, $\pm$25 and 0 microns). 

\begin{figure}
\includegraphics[width=\linewidth]{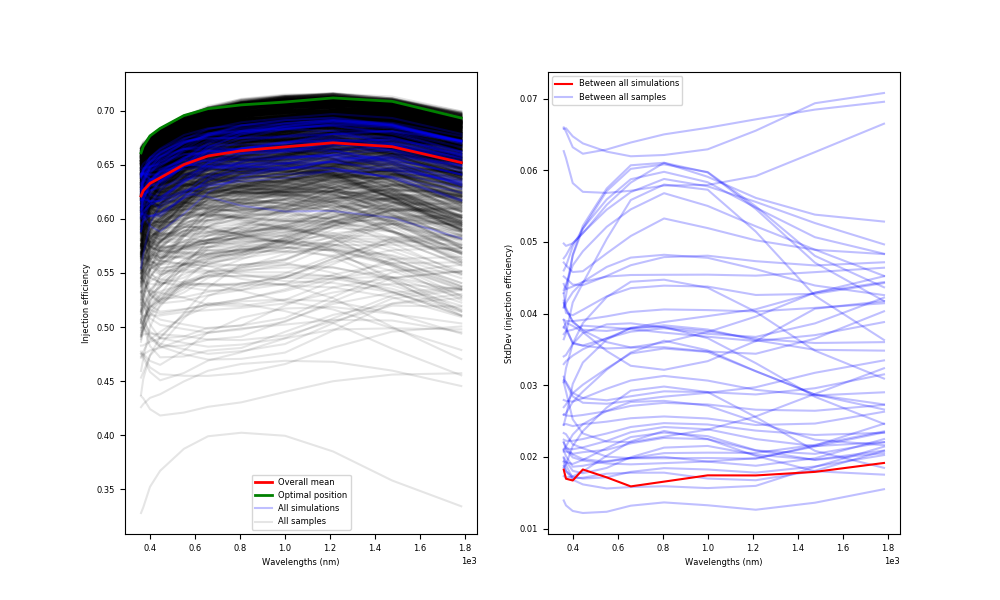}
\caption{\label{fig:ie_curve} IE curve for a fiber diameter of 1.0\arcsec, in a configuration at 30$\degree$ from Zenith, a field position 0.53$\degree$ from the optical axis, along the X-axis, a Moffat of 0.5\arcsec\ FWHM, and the weights for the LR spectrograph. The figure on the left show the IE curves: the red curve shows the average IE for the configuration, while the blue curves show the average IE for each simulation in that configuration, and the black curves show the IE for each sample. The green curve shows the IE at the optimal position. The figure on the right show the typical dispersion: in red the standard deviation between all simulations, and in blue the standard deviation between all samples in a given simulation.}
\end{figure}

We derive the IE for each simulation by averaging the injection efficiencies measured for all samples and then derive the IE for each configuration by averaging the injection efficiencies for each simulation. Figure \ref{fig:ie_curve} shows an example of the IE curves derived in a given configuration and the standard deviation measured between simulation and during a simulation (i.e. between samples). In the example shown, the standard deviation remains at about 2\% between simulations, and therefore below the 3\% spectrophotometric science requirements. It is however significantly larger, up to 10\%, between samples during a simulation.

The Figure also shows the IE curve at the optimal position, which is the best IE that can theoretically be reached. Some samples can sometimes be larger that the optimal IE curve, but only at some wavelengths, never overall. In the example shown, the simulated IE curve is about 5\% below the optimal IE curve

The IE curve for each configuration are then averaged over the field of view. Because the sampling over the field is coarse, with only twelve positions, we first interpolate the IE curves azimuthally and radially to populate the entire 1.5 square degree hexagonal field of view and then average them to derive a unique IE curve for each configuration (i.e. zenith distance, fiber diameter, and natural seeing).

Along with the current Throughput and Noise systems budgets, the Injection Efficiency systems budget answers the science requirements on sensitivity and together are used in the Exposure Time Calculator for MSE\footnote{\url{http://etc-dev.cfht.hawaii.edu/mse/}}.

\section{Conclusions}

In this paper we have presented a detailed budget for the IE of MSE, mostly relying on engineering information provided during conceptual design phase by the various subsystems design teams. We then used the Zemax model of the optical design for MSE with the systems budget for the IE to simulate the PSF at the focal plane and compute the IE in multiple configurations. We have shown that in the current configuration, the expected IE for MSE is only about 5\% below the optimal IE and that variations of the IE between two observations in the same configuration will remain smaller than 3\%.

\acknowledgements{The Maunakea Spectroscopic Explorer (MSE) conceptual design phase was conducted by the MSE Project Office, which is hosted by the Canada-France-Hawaii Telescope (CFHT). MSE partner organizations in Canada, France, Hawaii, Australia, China, India, and Spain all contributed to the conceptual design. The authors and the MSE collaboration recognize the cultural importance of the summit of Maunakea to a broad cross section of the Native Hawaiian community.}

\bibliography{report} 
\bibliographystyle{spiebib} 

\end{document}